\documentclass{aa}
\usepackage{graphicx, lscape}

\begin{document}
\title{Constraining the nature of High Frequency Peakers}
\subtitle{II. Polarization properties}
\author{
M. Orienti\inst{1,2,3} \and
D. Dallacasa\inst{2,3} 
}
\offprints{M. Orienti}
\institute{
Instituto de Astrofisica de Canarias, 38200-La Laguna, Tenerife, Spain \and
Dipartimento di Astronomia, Universit\`a di Bologna, via Ranzani 1,
I-40127, Bologna, Italy \and 
Istituto di Radioastronomia -- INAF, via Gobetti 101, I-40129, Bologna,
Italy 
}
\date{Received \today; accepted ?}

\abstract
{}
{The ``bright'' High Frequency Peakers (HFPs) sample is a mixture of
  blazars and intrinsically small and young radio sources. 
We investigate the polarimetric characteristics of 45 High
  Frequency Peakers, from the ``bright'' HFP sample, in order
  to have a deeper knowledge of the nature of each object, and to construct a
  sample made of
  genuine young radio sources only.}     
{Simultaneous VLA observations carried out at 22.2, 15.3, 8.4 and 5.0
  GHz, together with the information at 1.4 GHz provided by
  the NVSS at an earlier epoch,  
have been used to study the linearly polarized emission.}  
{From the analysis of the polarimetric properties of the 45 sources
we find that 26
  (58\%) are polarized at least at
  one frequency, while 
  17 (38\%) are completely unpolarized at all frequencies. We find a
  correlation between fractional polarization and the total intensity
  variability. We confirm that there is a clear distinction between
  the polarization properties of galaxies and quasars: 17 (66\%)
  quasars are highly polarized, while all the 9 galaxies are either
  unpolarized ($<0.2\%$) 
  or marginally polarized with fractional polarization below 1\%.
This suggests that
  most HFP candidates identified with quasars are likely to represent
a radio source
  population different from young radio objects.
}
{}
\titlerunning{Constraining the nature of HFP}
\maketitle
\section{Introduction}
From our previous works (Orienti et al. \cite{mo06};
Orienti, Dallacasa \& Stanghellini \cite{mo07}, hereafter Paper I) it
is clear that the bright HFP sample (Dallacasa et al. \cite{dd00})
contains both young radio sources and blazars, since it was
selected on the basis of single epoch multi-frequency radio
observations.
It is thus important to properly discriminate/classify these sources on the
basis of their intrinsic nature by using all the evidence coming
from observations.\\
The evolutionary stage of a powerful radio source in radio-loud
Active Galactic
Nuclei can be determined by its linear size. Following a self-similar
evolution model (e.g. Fanti et al. \cite{cf95}; Readhead et
al. \cite{read96}; Snellen et al. \cite{sn00}), the most compact
sources would evolve into the extended radio galaxy population and
eventually into the ``oldest'' and ``largest'' sources in
the Universe, the class of giant radio galaxies.\\
However, it has also been claimed (Alexander \cite{alexander00}; 
Marecki et al. \cite{ma03}) that 
a fraction of young and compact radio sources would die in an early
stage, never becoming large-scale objects.\\
In this scheme the youngest objects are the ``Compact Symmetric
Objects'' (CSOs), which are small ($<$ 1 kpc) radio sources with a
convex synchrotron radio spectrum which peaks at frequencies 
ranging from a few hundred MHz
to a few GHz (Wilkinson et al. \cite{wil94}).\\  
Statistical studies of different samples of this class of objects
(O'Dea \& Baum \cite{odea97}) have led to the discovery of an
anti-correlation between the spectral peak and the linear size
(i.e. the age): the higher the turnover frequency, the 
younger (smaller) the source is.\\
``High Frequency Peaker'' (HFP) radio sources (Dallacasa \cite{dd03}), 
characterized by the
same properties of the CSOs, but with observed spectral peaks above 5
GHz, are the best candidates to be newly born radio sources with
typical ages
of 10$^{2}$ - 10$^{3}$ years.\\
So far, the ``bright'' HFP sample (Dallacasa et al. \cite{dd00}) is
the only existing sample of this class of sources. Its selection
was based on radio spectral characteristics only.\\
Other kinds of radio sources, such as blazars, 
can temporarily match the selection criteria during a phase of 
their variability, 
i.e. when a flaring, self-absorbed component
at the jet base dominates the radio emission, and thus,
  they could contaminate
a sample selected on the basis of spectral properties.    
However, genuine young radio sources and blazars show different
characteristics, if proper observables are considered. 
The spectral variability (Tinti et al. \cite{st05};
Paper I) 
and the morphology (Orienti et
al. \cite{mo06}) of young HFP candidates 
have been discussed in earlier works.
This paper focuses on the polarimetric properties of these
two classes of objects.\\
Given their intrinsically small linear sizes, young radio sources
entirely reside within the Narrow Line Region (NLR) of the host galaxy.
The ambient medium of this region is generally described by a two-phase plasma:
a component in the 
form of clumps/clouds, with high density n$_{e}$ $\sim$ 10$^{4}$
cm$^{-3}$ and temperature T $\sim$ 10$^{4}$ K, but a small filling
factor (i.e. they occupy about $<$10$^{-4}$ of the total
volume of the NLR, McCarthy \cite{mac93}); 
and a diffuse, less dense 
(n$_{e}$ $\sim$ 10$^{-3}$ cm$^{-3}$), and hotter (T $\sim$ 10$^{7}$
K) component, which fills the inter-cloud space.
In the presence of a magnetic field, 
both components can act as a Faraday screen, causing significant
Faraday rotation 
and possibly depolarization of the synchrotron radiation.\\
Radio sources completely embedded in such an environment
are expected to show high Rotation measures (RM) and strong
depolarization (DP), if the structure of the Faraday Screen is not
resolved, which is generally the case, at least with arcsecond-scale
resolution polarimetric observations.\\
As the source expands ($>$ 1 kpc), its radio emission emerges from the
NLR and reaches the outer regions of the host galaxy interstellar
medium (ISM) where a more 
homogeneous
environment with less ionization and a weaker magnetic field 
may substantially reduce the aforementioned effects.\\
Evidence of a relationship between the fractional polarization and the
linear size has been pinpointed by Cotton et al. (\cite{cotton03}) and
Fanti et al. (\cite{cf04}) by studying samples of compact steep
  spectrum (CSS) and GHz-peaked spectrum (GPS) radio sources
spanning linear sizes from a fraction to a few kpc. In their work,
Cotton et al. (\cite{cotton03}) found that at 1.4 GHz almost all the
sources smaller than 6 kpc are completely unpolarized. At higher
frequencies (i.e. 5.0
and 8.4 GHz), Fanti et al. (\cite{cf04}) found a similar effect but the
complete depolarization of the radiation progressively happens at
smaller linear sizes ($<$ 3-5 kpc).\\   
Similar results have been obtained by Stanghellini et al. (\cite{cs98})
on a sample of GPS sources, whose typical linear size is $<$ 1
kpc. They found that the majority of the GPS objects have very low
fractional polarization, with upper limits 
consistent with the residual instrumental
polarization.\\   
On the other hand, blazars are usually large radio sources, but they appear
compact since their size is foreshortened by 
projection together with some amount of
beaming that enhances the emission of the core region and jet
base, 
making the large scale,
low-surface brightness emission barely visible in low dynamic range
observations.
In the unified scheme (e.g. Antonucci \cite{anto93}; Urry \& Padovani
\cite{urry95}), blazars are thought to be oriented at small angles to
the line of sight, allowing us to see their nuclear radiation directly and not
through the magneto-ionic medium and the obscuring torus. For this
reason they are expected to be significantly polarized, as is
usually
found (e.g. Saikia \cite{saikia99}). However, the high polarization
shown by these objects may also be due to Doppler boosted knots in
jets, as may be the case in 3C\,351 (Saikia \cite{saikia99}).\\
In this paper we present polarimetric data of
 45 HFP candidates from the ``bright'' HFP sample (Dallacasa et
 al. \cite{dd00}) observed with the VLA at frequencies from 4.6 to
 22 GHz. \\

\section{Polarization observations and data reduction}
Simultaneous multi-frequency observations of 45 (out of 55)
HFPs that were visible during the allocated observing time 
were carried out in July 2002 with the VLA in B configuration, 
in full polarization. 
The observing bandwidth was chosen to be 50 MHz per IF.
A separate analysis for each IF in L, C and X bands was carried out to
improve the spectral coverage of the data, as was done in previous works
(Dallacasa et al. \cite{dd00}; Tinti et al. \cite{st05}).\\
We obtained the flux density measurements 
in the L band (IFs at 1.465 and 1.665 GHz), C band (4.565 and
4.935 GHz), X band (8.085 and 8.465 GHz), U band (14.940 GHz) and K band
(22.460 GHz).\\
Each source typically was observed for 60 seconds in each band,
cycling through frequencies. Therefore, the flux density measurements
can be considered simultaneous.\\
Since the sources are relatively strong with small angular 
sizes, the snap-shot observing mode was considered adequate. 
For each frequency 
about 3 min were spent on each primary flux
density calibrator 3C\,286 and 3C\,48.
Secondary calibrators, chosen to minimize the telescope slewing time,  
were observed for 1.5 minutes at each frequency
every 25 minutes.
An appropriate calibrator (J1927+739) 
for the instrumental polarization was
observed over a wide range of parallactic angles.\\
The data reduction followed the standard procedures for the VLA,
implemented in the NRAO AIPS software.\\
After the standard amplitude and phase calibration, the instrumental
polarization was determined and removed.
The absolute orientation of the electric vector was
determined from the data of the primary flux density calibrator
3C\,286, which was observed twice. The residual instrumental
polarization is conservatively 
evaluated to be 0.1\%-0.3\%, and an uncertainty on the
orientation of about 2$^{\circ}$-3$^{\circ}$, depending on the
observing bands, being worse in U and K bands.\\ 
Total intensity data at all frequencies, together with the analysis of
the spectral variability have been published by
Tinti et al. (\cite{st05}).\\
Our measurements in the L band are less sensitive than those from the
NVSS (Condon et al. \cite{condon98}) due to
a less accurate calibration of the instrumental polarization,
mainly caused by some radio frequency interferences (RFI) 
which affected some scans of the calibrator
of the instrumental polarization. We then complemented our data with
the polarization measurements available from the NVSS, in order to
compare our sample with the results by Cotton et al. (\cite{cotton03})
and Fanti et al. (\cite{cf04}).\\
Polarization images  in the Stokes' U and Q parameters 
were produced for each frequency, with the 
exception of the L band. 

\begin{table*}[h!]
\tabcolsep 0.8mm
\begin{center}
\caption{Polarization properties.
Col. 1: source name (J2000); Col. 2: 
  optical identification (Table \ref{character})
 Col. 3: redshift; Col. 4: pc-scale
  morphology (see Table \ref{character}) 
Col. 5, 7, 9, 12, 15:
  polarization percentage $m$ and the error $\sigma_{m}$ 
in K, U, X, C and L bands
  respectively; Col. 6, 8, 10, 11, 13, 14, 16: polarization angle $\chi$
  and the error $\sigma_{\chi}$ 
in K, U, X, C and L bands. }
\begin{tabular}{cccc|cc|cc|ccc|ccc|cc}
\hline
Source&Id.&z&Morph.&m$^{K}$&$\chi ^{K}$&m$^{U}$&$\chi ^{U}$&m$^{X}$&$\chi
^{X}_{\rm 1}$&$\chi^{X}_{\rm 2}$&m$^{C}$&$\chi ^{C}_{\rm 1}$&$\chi ^{C}_{\rm 2}$&m$^{L}$&$\chi ^{L}$\\
(1)&(2)&(3)&(4)&(5)&(6)&(7)&(8)&(9)&(10)&(11)&(12)&(13)&(14)&(15)&(16)\\
\hline
&&&&&&&&&&&&&&&\\
J0003+2129&G&0.452&CSO&0.7$\pm$0.4&-35$\pm$11&0.6$\pm$0.4&16$\pm$10&
$<$0.2& - & - &$<$0.2& - & - &$<$0.1& -\\
J0005+0524&Q&1.887&CSO&-& - &- & -
&0.5$\pm$0.3&15$\pm$4&37$\pm$6&$<$0.2& - & - &$<$0.3& - \\
{\bf J0037+0808}&EF& &CSO&$<$0.3& - &$<$0.3& - &$<$0.2& - & - &$<$0.2& -
& - &$<$0.5& -\\
{\bf J0111+3906}&G&0.668&CSO&$<$0.1& - &$<$0.1& -&$<$0.1&
-&-&$<$0.1& - & - &$<$0.1&-\\
{\bf J0116+2422}&EF& &Un&$<$0.3& -& -& -& $<$0.2& - & - &$<$0.2& - &- &0.4&-\\
J0217+0144&Q&1.715&Un&1.0$\pm$0.1&71$\pm$3&2.2$\pm$0.1&73$\pm$3&2.7$\pm$0.1&87$\pm$2&87.3$\pm$2
&1.7$\pm$0.1&83$\pm$2&90$\pm$2&1.5$\pm$0.1&42$\pm$1.0\\
J0329+3510&Q&0.5&CJ&1.6$\pm$0.2&-1$\pm$3&1.6$\pm$0.1&35$\pm$3&0.5$\pm$0.1&59$\pm$3&44$\pm$3&1.0$\pm$0.1&-75$\pm$2&-73$\pm$2&5.0$\pm$0.3&-46$\pm$1\\
J0357+2319&Q& &Un&3.2$\pm$0.3&88$\pm$3&2.2$\pm$0.3&-87$\pm$4&1.7$\pm$0.4&-80$\pm$3&-78$\pm$3&1.5$\pm$0.4&-68$\pm$3&-80$\pm$3&0.8$\pm$0.3&-16$\pm$6\\
{\bf J0428+3259}&G&0.479&CSO&$<$0.2&-&$<$0.1&-&$<$0.1&-&-&$<$0.1&
- & - &$<$0.3& - \\
J0519+0848&EF& &Un&1.6$\pm$0.2&55$\pm$3&-&-&1.0$\pm$0.1&
-89$\pm$2&89$\pm$3&2.0$\pm$0.1&86$\pm$2&85$\pm$2&2.3$\pm$0.3&-6$\pm$2\\
J0625+4440&BL& &Un&0.7$\pm$0.2&-75$\pm$5&-&-&2.1$\pm$0.2&-87$\pm$3&-86$\pm$2&2.4$\pm$0.2&-86$\pm$3&-86$\pm$2&4.7$\pm$0.4&-79$\pm$1\\
{\bf J0638+5933}&EF& &CSO&$<$0.1&-&-&-&$<$0.1& - & - &$<$0.1&
- & - &$<$0.2& - \\
J0642+6758&Q&3.180&Un&$<$0.3&-&0.8$\pm$0.2&-43$\pm$6&0.6$\pm$0.1&-69$\pm$3&-60$\pm$3&1.6$\pm$0.1&-81$\pm$2&-84$\pm$2&$<$0.2& - \\
J0646+4451&Q&3.396&MR&2.5$\pm$0.2&19$\pm$3&3.6$\pm$0.1&17$\pm$3&3.1$\pm$0.1&30$\pm$2&28$\pm$2&1.6$\pm$0.1&29$\pm$2&31$\pm$2&0.8$\pm$0.1&44$\pm$2\\
{\bf J0650+6001}&Q&0.455&CSO&$<$0.1&-&-&-&$<$0.1& - & - &$<$0.1& - & -
&$<$0.1& - \\
{\bf J1335+4542}&Q&2.449&CSO&$<$0.2&-&-&-&$<$0.1&-& - &$<$0.1&-& - &$<$0.2&-\\
{\bf J1335+5844}&EF&-&CSO&$<$0.2&-&-&-&$<$0.1&-& - &$<$0.1&-& -
&$<$0.1& -\\
{\bf J1407+2827}&G&0.0769&CSO&$<0.2$&-&-&-&$<$0.2& - & -
&$<$0.1&-& - &$<$0.1&-\\
{\bf J1412+1334}&EF& &Un&$<$0.3&-&-&-&$<$0.2&-& - &$<$0.1& -
& - &$<$0.4& - \\
J1424+2256&Q&3.626&Un&-&-&-&-&3.4$\pm$0.1&3$\pm$2&3$\pm$2&0.4$\pm$0.1&62$\pm$3&-89$\pm$3&0.6$\pm$0.2&3$\pm$5\\
J1430+1043&Q&1.710&MR&-&-&-&-&$<$0.1&-& - &$<$0.1& - & - &0.5$\pm$0.2&-71$\pm$6\\
J1457+0749&BL& &Un&-&-&2.8$\pm$0.3&-39$\pm$4&5.3$\pm$0.1&-35$\pm$2&-37$\pm$2&5.7$\pm$0.2&-33$\pm$2&-34$\pm$2&- & - \\
J1505+0326&Q&0.411&Un&1.1$\pm$0.1&-43$\pm$3&-&-&0.7$\pm$0.1&-45$\pm$3&-44$\pm$3&0.2$\pm$0.1&-41$\pm$3&-45$\pm$3&1.6$\pm$0.1&56$\pm$1\\
{\bf J1511+0518}&G&0.084&CSO&$<$0.1&-&-&-&$<$0.1&-& -
&$<$0.1& - & - &$<$0.7& - \\
{\bf J1526+6650}&Q&3.02&MR&$<$0.4&-&$<$0.3&-&$<$0.1& - & - &$<$0.1& - & - &$<$0.6&-\\
J1603+1105&BL& &Un&$<$0.2&-&$<$0.3&-&0.9$\pm$0.2&-39$\pm$3&-36$\pm$3&1.8$\pm$0.2&-3$\pm$3&-6$\pm$3& - & - \\
J1616+0459&Q&3.197&CJ&2.3$\pm$0.2&90$\pm$4&-&-&2.9$\pm$0.1&-86$\pm$3&84$\pm$3&0.4$\pm$0.1&-41$\pm$3&-66$\pm$3&$<$0.2&-\\
{\bf J1623+6624}&G&0.203&Un&$<$0.3&-&$<$0.2&-&$<$0.2& - & - &$<$0.2& -& - &$<$0.3&-\\
J1645+6330&Q&2.379&Un&2.9$\pm$0.3&78$\pm$3&2.7$\pm$0.1&73$\pm$3&3.0$\pm$0.1&63$\pm$3&64$\pm$3&1.3$\pm$0.1&46$\pm$2&45$\pm$2&2.4$\pm$0.2&-15$\pm$2\\
{\bf J1735+5049}&G&
&CSO&$<$0.1&-&$<$0.1&-&$<$0.1&-&- &$<$0.1&
- & - &$<$0.1&-\\
J1800+3848&Q&2.092&Un&0.6$\pm$0.1&-84$\pm$3&-&-&0.3$\pm$0.1&-62$\pm$3&-24$\pm$3&$<$0.1&
- & - &0.6$\pm$0.1&56$\pm$5\\
J1811+1704&BL& &CJ&7.1$\pm$0.7&-50$\pm$3&6.7$\pm$0.3&-42$\pm$3&1.8$\pm$0.1&12$\pm$3&9$\pm$3&2.3$\pm$0.1&-19$\pm$3&-27$\pm$3& -& - \\
J1840+3900&Q&3.095&Un&1.0$\pm$0.4&85$\pm$8&-&-&$<$0.3&-&- &$<$0.3&
- & - &$<$0.4& - \\
{\bf J1850+2825}&Q&2.560&MR&$<$0.1&-&$<$0.1&-&$<$0.1& - & - &$<$0.1& -
& - &$<$0.3& - \\
{\bf J1855+3742}&G& &CSO&$<$0.4&-&$<$0.4&-&$<$0.2& - & - &$<$0.1& - &
- &$<$0.3& - \\
J2021+0515&Q&
&CJ&1.6$\pm$0.2&47$\pm$4&1.4$\pm$0.2&-76$\pm$4&$<$0.1& -
& - &$<$0.3$\pm$0.1&- & - &$<$0.3& - \\
J2024+1718&Q&1.05&Un&0.9$\pm$0.1&45$\pm$3&0.6$\pm$0.1&-32$\pm$4&0.8$\pm$0.1&45$\pm$3&45$\pm$3&0.2$\pm$0.1&
-2$\pm$3&-5$\pm$3&$<$0.3&-\\
J2101+0341&Q&1.013&Un&2.5$\pm$0.1&-68$\pm$3&3.5$\pm$0.1&-69$\pm$3&4.5$\pm$0.1&-70$\pm$2&-68$\pm$2&3.7$\pm$0.1&-58$\pm$2&-60$\pm$2&$<$0.1&
- \\
J2123+0535&Q&1.878&CJ&3.7$\pm$0.2&-1$\pm$3&2.7$\pm$0.1&1$\pm$3&0.8$\pm$0.1&22$\pm$3&15$\pm$3&1.1$\pm$0.1&82$\pm$3&82$\pm$3&2.0$\pm$0.1&53$\pm$1\\ 
{\bf J2203+1007}&G& &CSO&$<$0.4&-&$<$0.4&-&$<$0.2& - & - &$<$0.2& - &
- &$<$0.4& - \\
J2207+1652&Q&1.64&Un&$<$0.3&-&$<$0.3&-&1.5$\pm$0.2&85$\pm$3&87$\pm$3&1.3$\pm$0.2&-72$\pm$3&-80$\pm$3&3.6$\pm$0.2&55$\pm$1\\
J2212+2355&Q&1.125&Un&8.6$\pm$0.4&36$\pm$3&7.9$\pm$0.4&38$\pm$3&6.8$\pm$0.2&28$\pm$2&28$\pm$2&5.7$\pm$0.2&-4$\pm$2&1$\pm$2&2.0$\pm$0.1&43$\pm$1\\
J2257+0243&Q&2.081&Un&$<$0.2&-&$<$0.2&-&$<$0.3&-&-&$<$0.1&-&
- &0.9$\pm$0.2&44$\pm$1\\
J2320+0513&Q&0.622&Un&0.9$\pm$0.1&69$\pm$3&1.3$\pm$0.1&44$\pm$3&2.7$\pm$0.1&59$\pm$2&58$\pm$2&4.3$\pm$0.1&71$\pm$2&71$\pm$2&2.3$\pm$0.1&59$\pm$1\\
J2330+3348&Q&1.809&MR&1.1$\pm$0.1&44$\pm$3&$<$0.2&-&0.5$\pm$0.1&-33$\pm$3&-33$\pm$3&1.4$\pm$0.1&-17$\pm$3&-15$\pm$3&1.2$\pm$0.2&-39$\pm$4\\ 
&&&&&&&&&&&&&&&\\
\hline
\end{tabular} 
\end{center}
\label{provola}
\end{table*}  

The final noise (1$\sigma$) is usually of
$\sim$ 0.07 mJy/beam for C and X bands, and $\sim$ 0.2 
  mJy/beam for U and K
bands, although for a few objects a higher noise level was
  found in these latter bands, limiting the sensitivity in 
detect-polarized emission.\\
Images of the polarization intensity, polarization angle and
percentage polarization were produced for all the sources.\\
Based on the Stokes' parameters and their errors the polarization
parameters were calculated for each frequency as in Klein et
al. (\cite{uk03}).\\ 
Polarization
percentage and polarization angle with their errors 
at each frequency are reported
in Table 1. In C and X bands we report the polarization angle for each
single frequency. Data concerning the L band are from the NVSS (Condon
et al. \cite{condon98}). 
Polarization angles represent the position angle of the
E-vector as measured by integrating over each source (AIPS tasks JMFIT
and IMSTAT) on the Stokes U and Q images.
Sources marked in boldface are those found
unpolarized (m $< 0.5\%$) at all frequencies.
Table \ref{character} summarizes the optical
identification and the radio structure of the observed sources.\\
In the case of C and X bands, where two frequencies per band are
available, Table 1 provides only one value for the
polarization percentage since it does not change significantly between
the frequencies. Conversely, for the polarization angle we prefer to
report the values for each separate frequency. 
For unpolarized sources i.e. those with a 
signal-to-noise ratio below 3$\sigma$ in polarized emission, 
the polarization angle is not provided.\\  

\begin{table}
\begin{center}
\caption{Summary of optical identification versus radio structure
  (Orienti et al. \cite{mo06})
  of
  the 45 observed sources.} 
\begin{tabular}{|c|c|c|c|c|c|}
\hline
 &G&Q&BL&EF&Tot\\
\hline
&&&&&\\
CSO& 8& 3& 0& 3&14\\
CJ & 0& 4& 1& 0& 5\\
Un & 1&14& 3& 3&21\\
MR & 0& 5& 0& 0& 5\\
Tot& 9&26& 4& 6&45\\
&&&&&\\
\hline
\end{tabular}  
\label{character}
\end{center}
\end{table}

\section{Results}
The intrinsic radio polarization, the rotation measure and the
depolarization are discriminant ingredients in the determination of
different classes of extragalactic radio sources.\\
Our multi-frequency 
VLA polarization measurements, in addition to the information
from the NVSS allow us to identify and remove contaminating
objects from the sample of candidate young HFPs
(Dallacasa et al. \cite{dd00}).\\

\subsection{Fractional polarization} 

The properties of the polarized emission of young radio sources and
blazars are very different. 
As previously mentioned, statistical studies (e.g. Fanti et
al. \cite{cf04}) show that young radio sources
have a monotonic decrement in their fractional polarization with
decreasing frequency, consistent with substantial depolarization. 
In contrast, blazars show a relatively constant polarization
at all frequencies (Klein et al. \cite{uk03}).\\
Considering the VLA data (from K to L band), we find that 12 ($\sim 26\%$)
sources (Table 1)
have significant polarized emission  
at all our frequencies: 9  
of them with $m$ greater than about 1\% 
(6 quasars J0217+0144, J0357+2319, J0646+4451, J1645+6330,
J2212+2355 and J2320+0513, 2 BL Lacs J1457+0749 and
J1811+1704 and the Empty Field J0519+0848); 
a further 3 objects (2 quasars J0329+3510 and J2123+0535 and the BL Lac
J0625+4440) have well detected polarized emission at all the available
frequencies but the fractional polarization may drop below 1\%.\\ 
On the other hand, 17 ($\sim 38\%$) 
sources (marked in boldface in Table 1)
are completely
unpolarized at any frequency. 
Of the remaining sources, 11 ($\sim24\%$) objects (10 quasars
J0642+6758, J1424+2256, J1505+0326, J1616+0459, J1840+3900,
J2021+0515, J2101+0341, J2207+1652, J2257+0243 
and J2330+3348 and the BL Lac J1603+1105)
have irregular fractional polarization with values as high as 3\% in
one band, while its linear polarization may be undetected at other
(either higher or lower) frequencies (either higher or lower).
5 ($\sim11\%$) objects
(1
galaxy J0003+2129 and 4 quasars J0005+0524, J1430+1043,
J1800+3848 and J2024+1718)
are slightly polarized ($m > 0.5\%$) at least at one frequency. \\
The comparison between polarized properties and the flux density
variability can be more effective in determining the nature of
candidate HFP. 
Indeed, blazar objects 
display 
significant  polarization, as well as 
a high degree of flux density variability. 
Conversely, young radio sources are the least variable class of
extragalactic objects (O'Dea \cite{odea98}) with a mean variation within
$\sim$5\% (Stanghellini et al. \cite{cstan05}).\\
We thus compare the fractional polarization $m$ at each frequency 
with the total intensity flux density variability V, as described in 
Paper I.
Sources with variability index $V$$<$3 are characterized by the lack
of flux density variability and are good candidates to be genuine
young radio sources, while sources with $V$$>$3 do show
significant variability and they are likely part of the blazar population.\\
In order to investigate a possible separation in the polarimetric
properties between sources with different variability index, 
we compare the median $m$ at each frequency for {\it all}
the observed sources and for those with V$<$ 3 and V$>$ 3 (Table
\ref{fractio}).\\

\begin{table*}
\begin{center}
\caption{Median degrees of percentage polarization and the statistical 
errors.  
L-band data are from the NVSS (Condon et
  al. \cite{condon98}). }
\begin{tabular}{|c|l|l|l|l|l|l|l|}
\hline
 &$\;\;\;\;\;\;\;$All&$\;\;\;\;\;\;$V$<$3 &$\;\;\;\;\;\;$V$>$3&$\;\;\;\;\;\;$G&$\;\;\;\;\;\;$Q&$\;\;\;\;\;\;$BL&$\;\;\;\;\;\;$EF \\
Band&$\overline{m}\;\;\;\;\;\;\;\;$$\sigma_{\overline{m}}$&$\overline{m}\;\;\;\;\;\;\;\;$$\sigma_{\overline{m}}$&$\overline{m}\;\;\;\;\;\;\;\;$$\sigma_{\overline{m}}$&$\overline{m}\;\;\;\;\;\;\;\;$$\sigma_{\overline{m}}$&$\overline{m}\;\;\;\;\;\;\;\;$$\sigma_{\overline{m}}$&$\overline{m}\;\;\;\;\;\;\;\;$$\sigma_{\overline{m}}$&$\overline{m}\;\;\;\;\;\;\;\;$$\sigma_{\overline{m}}$\\
\hline
&&&&&&&\\
L&0.40$\;\;\;\;\;$0.18&0.30$\;\;\;\;\;$0.05&0.80$\;\;\;\;\;$0.27&0.30$\;\;\;\;\;$0.07&0.60$\;\;\;\;\;$0.23&-$\;\;\;\;\;\;\;\;\;\;$-&0.40$\;\;\;\;\;$0.34\\
C&0.20$\;\;\;\;\;$0.21&0.10$\;\;\;\;\;$0.01&1.30$\;\;\;\;\;$0.29&0.10$\;\;\;\;\;$0.02&0.40$\;\;\;\;\;$0.28&2.35$\;\;\;\;\;$0.89&0.15$\;\;\;\;\;$0.31\\
X&0.30$\;\;\;\;\;$0.23&0.15$\;\;\;\;\;$0.03&0.90$\;\;\;\;\;$0.32&0.20$\;\;\;\;\;$0.02&1.65$\;\;\;\;\;$0.33&1.95$\;\;\;\;\;$0.96&0.20$\;\;\;\;\;$0.14\\
U&0.60$\;\;\;\;\;$0.37&0.30$\;\;\;\;\;$0.06&1.50$\;\;\;\;\;$0.48&0.20$\;\;\;\;\;$0.07&1.40$\;\;\;\;\;$0.47&2.80$\;\;\;\;\;$1.86&-$\;\;\;\;\;\;\;\;\;\;$-\\
K&0.40$\;\;\;\;\;$0.28&0.25$\;\;\;\;\;$0.05&1.00$\;\;\;\;\;$0.40&0.20$\;\;\;\;\;$0.07&1.00$\;\;\;\;\;$0.39&0.7$\;\;\;\;\;$2.22&0.30$\;\;\;\;\;$0.23\\
&&&&&&&\\
\hline
\end{tabular}
\end{center}
\label{fractio}
\end{table*}

\begin{figure*}
\begin{center}
\includegraphics{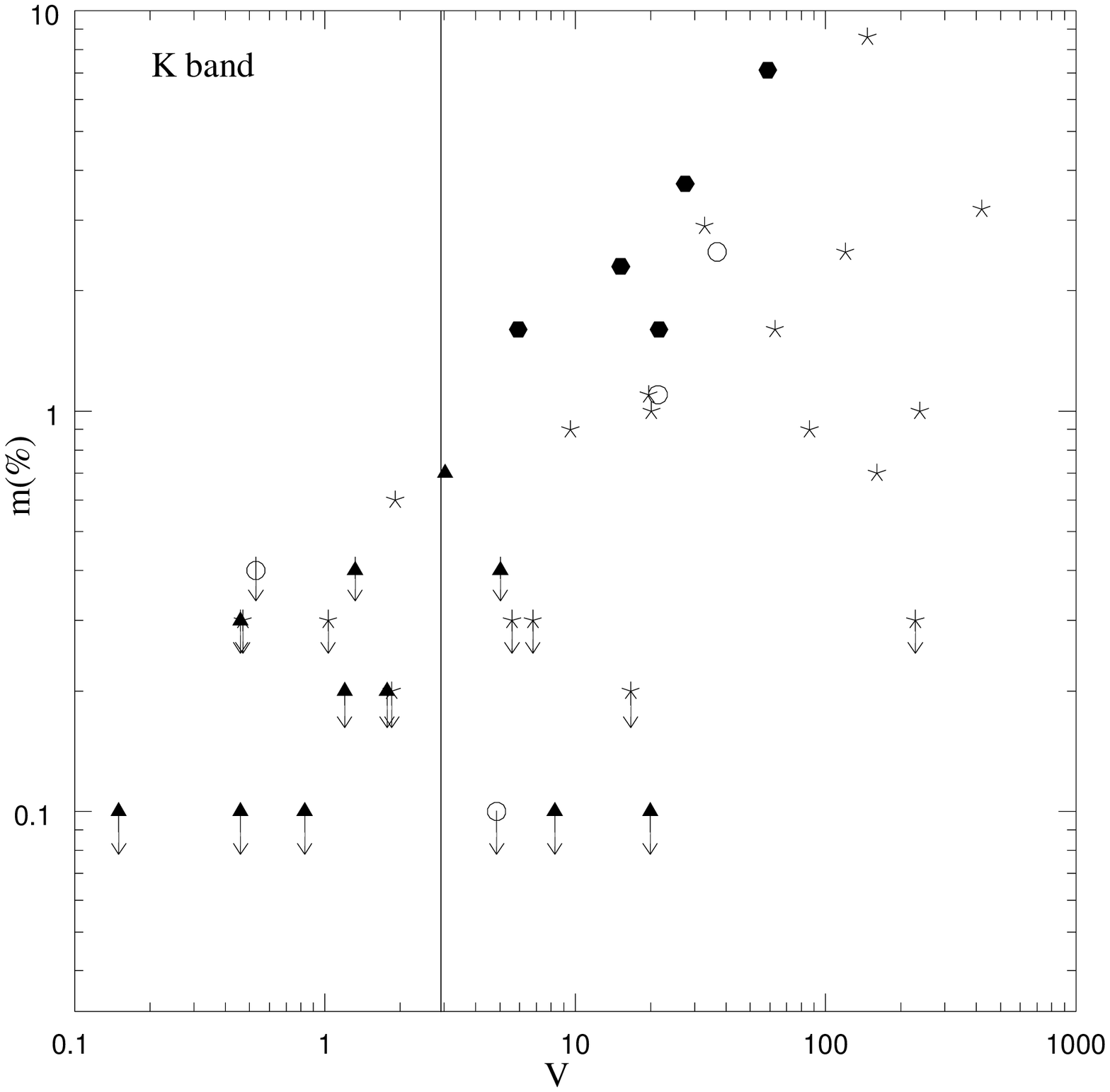}
\includegraphics{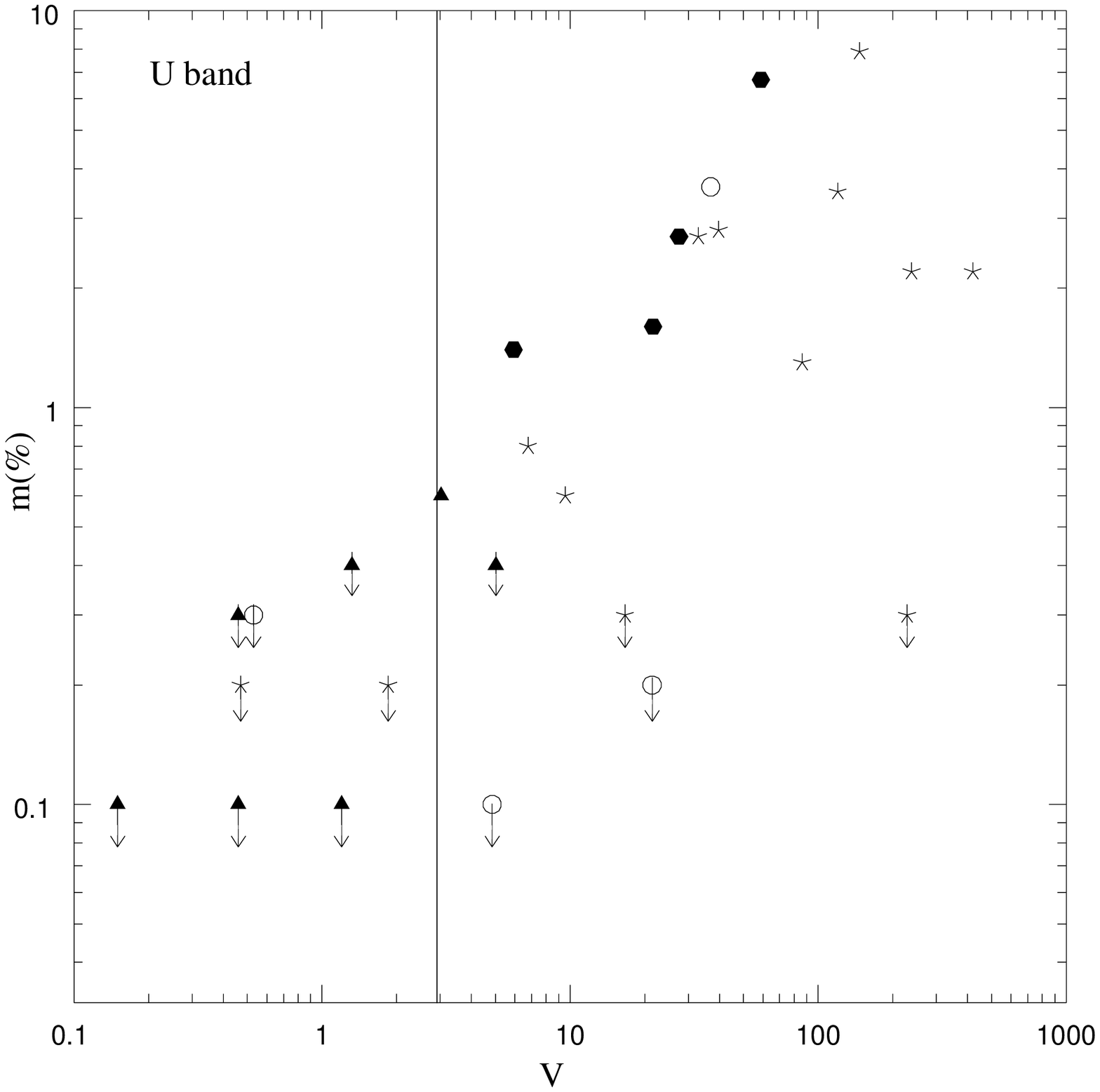}
\includegraphics{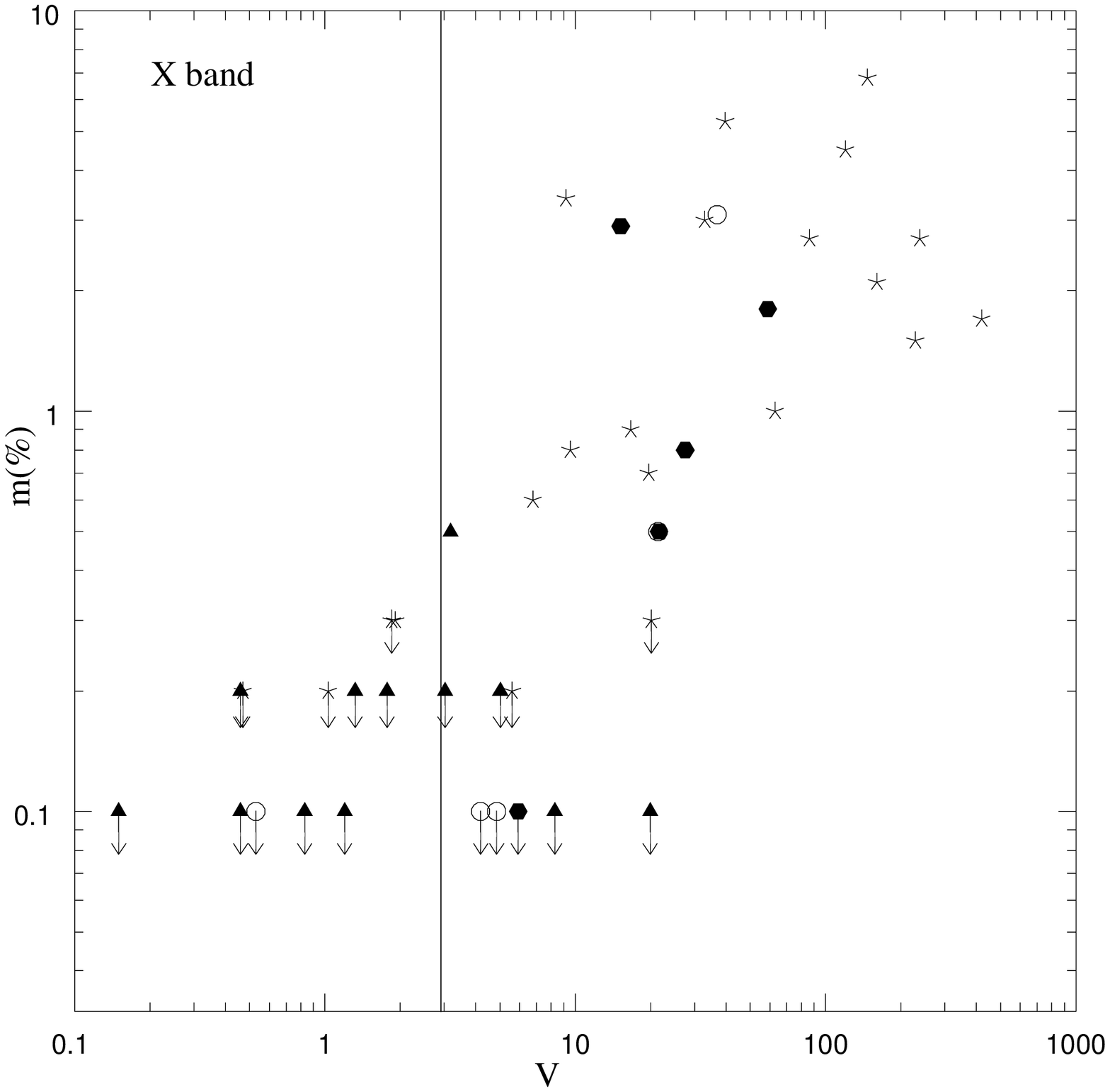}
\includegraphics{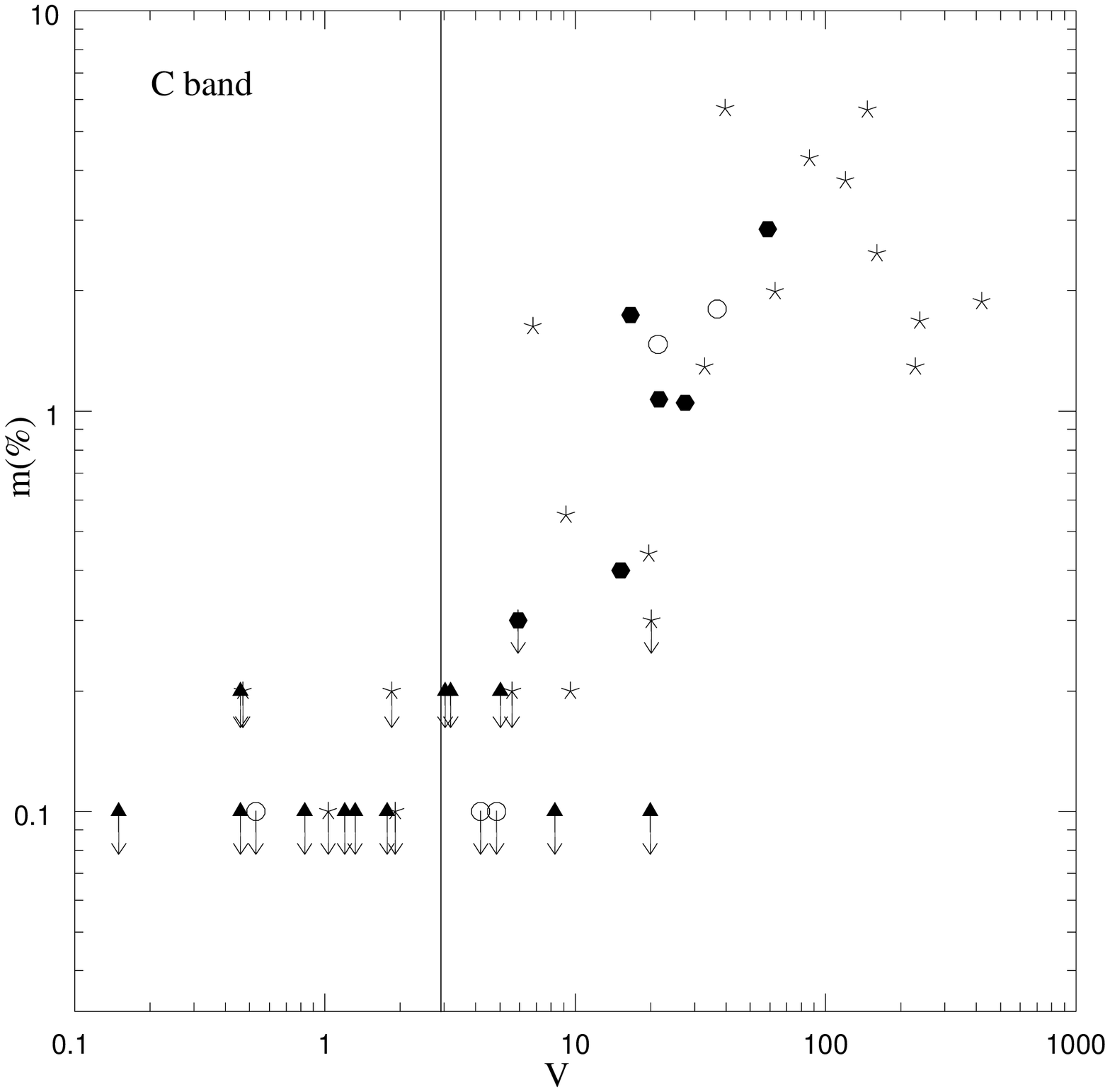}
\includegraphics{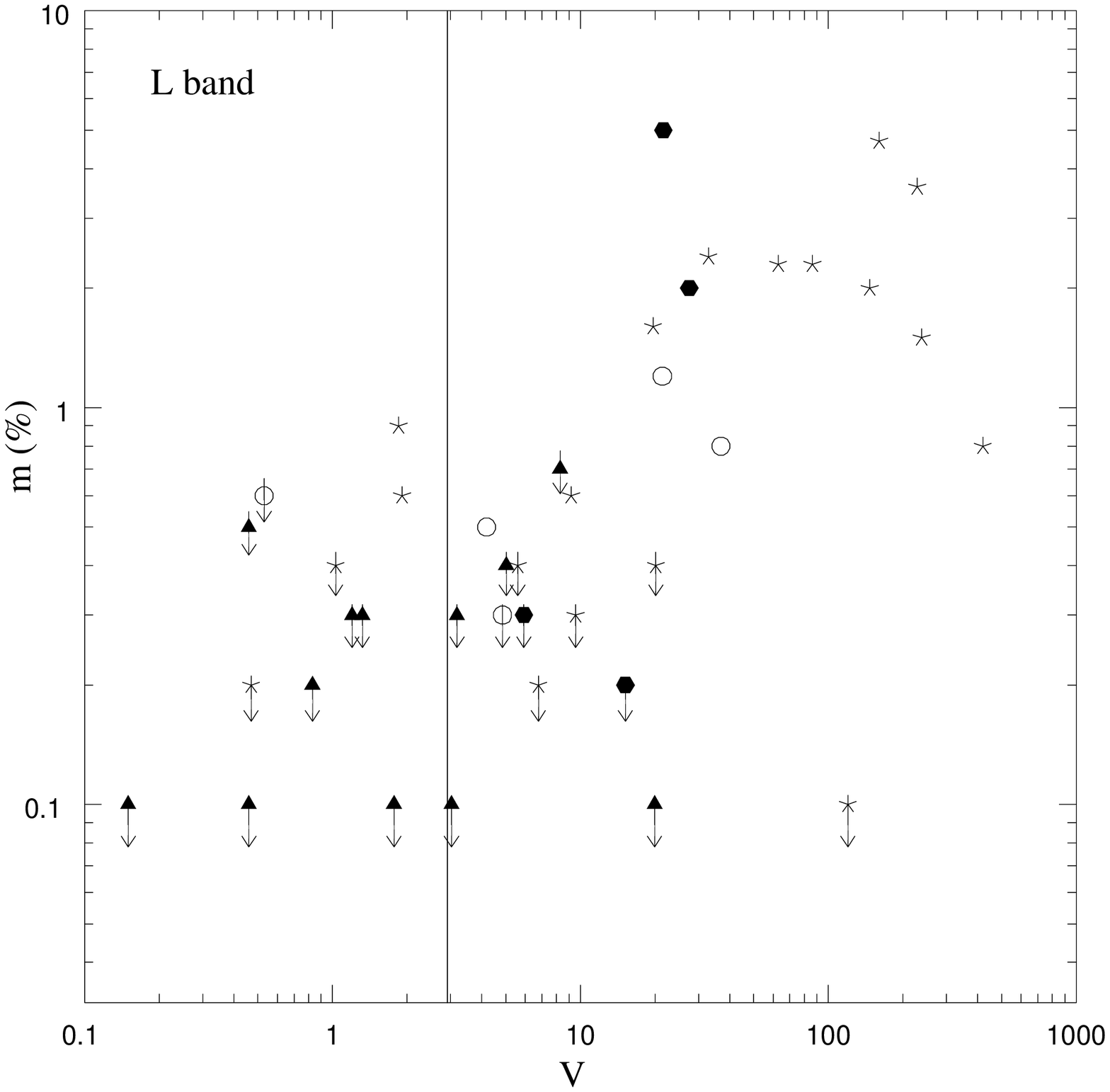}
\vspace{14.5cm}
\caption{Polarization percentage in K, U, X, C and L bands. 
L-band data are from the NVSS (Condon et al. \cite{condon98}).
The vertical
  line represents the total intensity flux density variability
  index $V=3$ (Paper I). }
\label{plot}
\end{center}
\end{figure*}

There is a well established difference in the polarization percentage
at all frequencies between
sources with different variability index:
17 objects (94\%) 
of the sources with $m>\, \sim 1\%$ at at least one frequency 
are strongly variable ($V\gg3$)
and among them 11 objects ($\sim$65\%) 
no longer show the convex radio spectrum (Paper I), 
while 14 objects ($\sim 82\%$) of the
unpolarized sources have $V<3$.\\
In Fig. \ref{plot} plots of the polarization percentage versus
the total intensity flux density variability in K, U, X, C and L bands
are shown. Filled triangles, filled circles, empty circles 
and stars represent sources
  with a CSO structure, a Core-Jet morphology, an unresolved and a
 marginal structure respectively. Upper limits are indicated with 
an arrow associated with each symbol.\\
 
From these diagrams it is clear that significant fractional
polarization is found only in sources with $V >3$.  
The median $V$ of
the sources with high $m$ (i.e. larger than the median $m$ of
the sample) is about 10 times higher than that of sources with
low $m$, independent of frequency.
Conversely, the median $m$ of the sources with $V>3$ is higher than the median
$m$ of sources $V<3$ at all frequencies (Table \ref{fractio}). 
If we consider the fractional polarization in relation to the
pc-scale structure (Table 1), we find that 
12 objects (86\%) of the sources
with a CSO-like morphology (represented by filled triangles in
Fig. \ref{plot}) are unpolarized at all frequencies, while 
26 objects (84\%) of
the sources without a CSO-like structure
are polarized at at least one frequency. 
Only two sources with a CSO-like morphology (the galaxy
J0003+2129 and the quasar J0005+0524) 
do show some amount of
polarized emission ($m>0.5\%$) at the highest frequencies, while
they are unpolarized in the C band, as expected if an unresolved
Faraday Screen is present along the line of sight.\\  
The comparison between the fractional polarization and the pc-scale
structure (Orienti et al. \cite{mo06}) suggests that
sources with or without a CSO-like morphology, typical of very young radio
sources, have a different degree of
polarization.\\  
Although the L-band data are from a different epoch, 
we find the same trend of fractional polarization as in higher
frequency data. Two quasars (J1430+1043 and J2257+0243) show
polarized emission in the L-band only.
This suggests
that, except in these two cases, 
the polarized emission mainly originates in the central region of radio
sources without significant contribution from extended features.\\
To investigate whether radio sources with different optical
counterparts have different polarimetric properties, we compare the
median polarization percentage of sources associated with galaxies, quasars,
BL Lacs and empty fields (Table \ref{fractio}). Although there
  are limited
statistics available, we find that galaxies and empty fields have very
low values at all frequencies, while quasars and BL Lacs are more
polarized, as expected in unified schemes. In this scheme radio
galaxies lie close to the plane of the sky, while quasars and BL Lacs
are observed at small viewing angles. Thus the radio emission from the
central region of quasars and BL Lacs undergoes little influence
from the
magneto-ionic medium and the obscuring torus/disc, as happens in 
galaxies, and little or even no 
depolarization is expected. No relationship between
the fractional polarization and the observing frequencies has been
found, but this is likely due to the poor statistics at high frequency.\\

\subsection{Depolarization and rotation measure}

The basic theory of the Faraday effects on polarized emission is
described in Burn (\cite{burn66}).
In the following discussion we assume that the radiation leaving the
volume occupied by the magnetic field and the relativistic particles
is polarized at some level. Then, it crosses a region where it
undergoes Faraday rotation and depolarization. These phenomena are
more severe where the ion (electron) density is high and there is a
disordered magnetic field with a substantial component along the line
of sight.\\
We have used our measurements of depolarization,

\begin{displaymath}
DP= {\large \frac{m_{\nu1}}{m_{\nu2}}}, \, \,\,\,\, {\rm with}\,\, \nu_{1} < \nu_{2}
\end{displaymath}

\noindent and of $\chi_{\nu}$ (orientation of the electric
polarization vector at a given frequency $\nu$) to derive the rotation
measure\\

\begin{displaymath}
RM = 0.81 \int_{L} n_{e}B_{\parallel}dL\; (\rm rad\;m^{-2})
\end{displaymath}

\noindent where $n_{e}$ is the electron density of thermal plasma in
cm$^{-3}$,
$B_{\parallel}$ the magnetic field component along the line of sight in
$\mu$G and L the effective path length in parsecs.
In general, multiple frequencies are necessary to solve for the $n\pi$
ambiguities in measurements.\\
From multi-frequency observations, RM can be determined 
by the well known relation\\

\begin{displaymath}
\chi_{\rm obs} = \chi_{\rm int}+RM \cdot \lambda^{2}
\end{displaymath}

\noindent where the intrinsic position angle $\chi_{\rm int}$ is the
zero-wavelength value.\\
In general the $\lambda^{2}$-law comes from the assumptions of the
ideal model as in Burn (\cite{burn66}). Deviations from the
$\lambda^{2}$-law may arise when the total polarization is the
superposition of two or more regions with different $m$ and $RM$ (see
e.g. Rossetti et al., submitted).\\ 
For the depolarization we consider data in K, X, C and L 
bands since
in U band the availability of only a few polarization measurements
does not allow an accurate statistical analysis.\\ 
We compute the depolarization DP for all the sources with substantial
polarized emission (only upper limits at the lowest frequency have
been included; Table 1). 
The median values found are consistent with a slight
depolarization (Table \ref{tab-dep}), with the exception of
DP$^{\rm L}_{\rm C}$, where it seems that L-band data are more polarized
than C-band data. This may happen when polarization in L-band comes
from extended features that are resolved out in the C-band and 
when contributions from the core region are still negligible.\\

\begin{table}
\begin{center}
\caption{Median degrees of depolarization and statistical
  errors for the sources in which
  polarized emission has been detected. Only sources with upper limits at the
  lower frequency (Table 1) have been
  considered.}
\begin{tabular}{|c|l|}
\hline
 &$\overline{DP}\;\;\;\;\;\;\;\;$$\sigma_{\overline{DP}}$\\
\hline
&\\
L/C&1.15$\;\;\;\;\;\;$0.45\\
L/X&0.80$\;\;\;\;\;\;$0.52\\
L/K&0.69$\;\;\;\;\;\;$0.42\\
C/X&0.87$\;\;\;\;\;\;$0.18\\
C/K&0.55$\;\;\;\;\;\;$0.27\\
X/K&0.72$\;\;\;\;\;\;$0.22\\
&\\
\hline
\end{tabular}
\end{center}
\label{tab-dep}
\end{table}

\begin{figure}
\begin{center}
\includegraphics{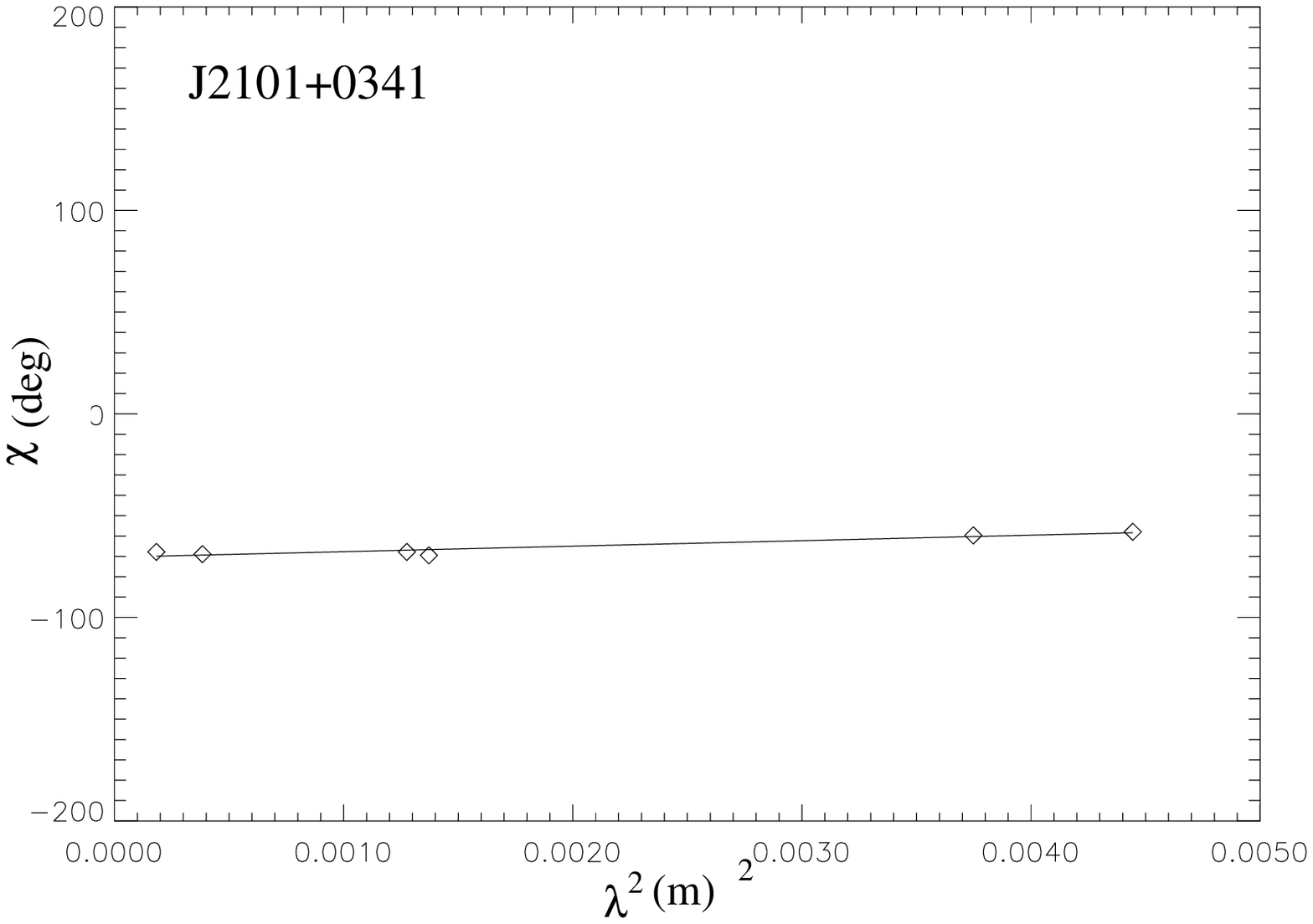}
\includegraphics{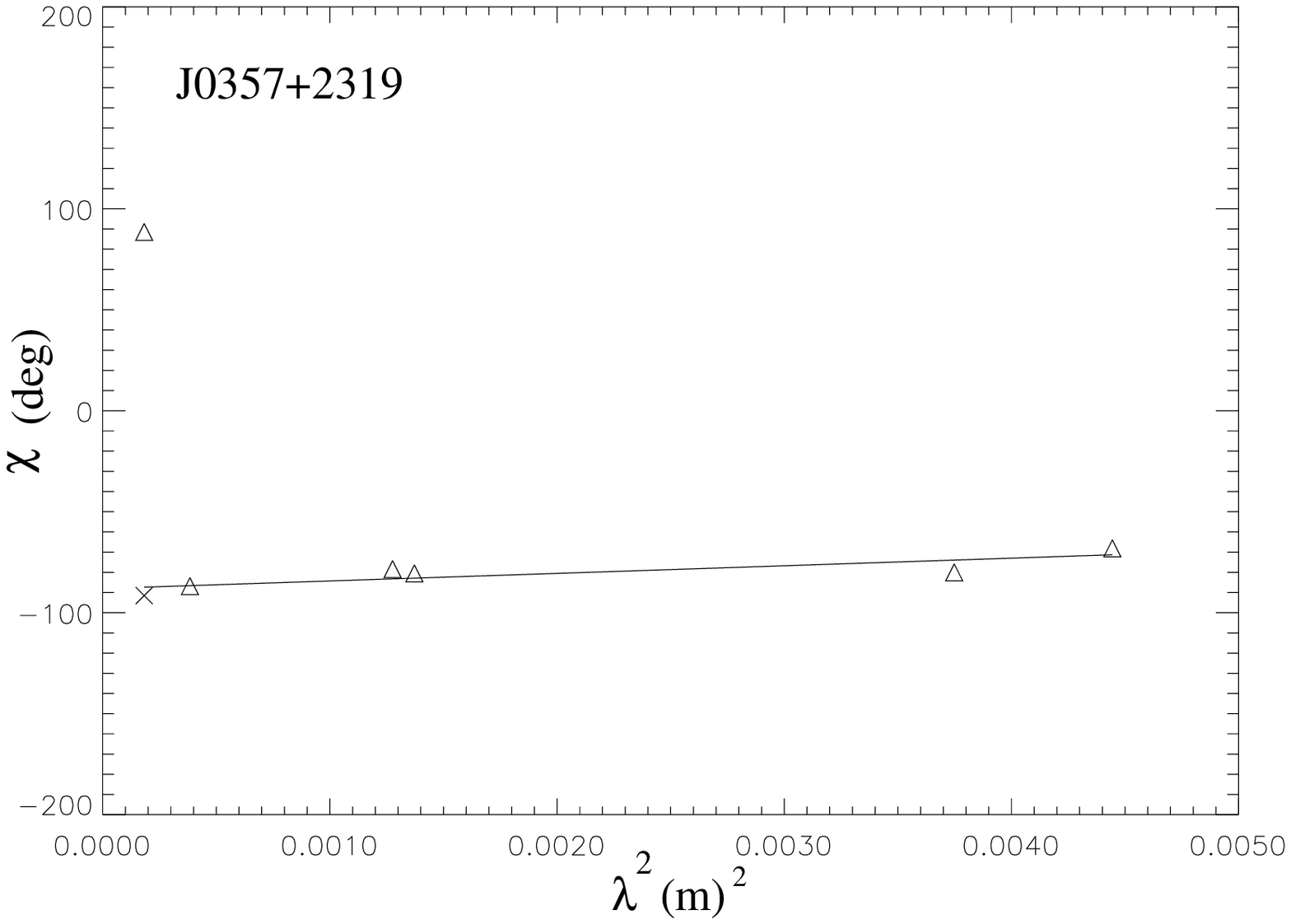}
\includegraphics{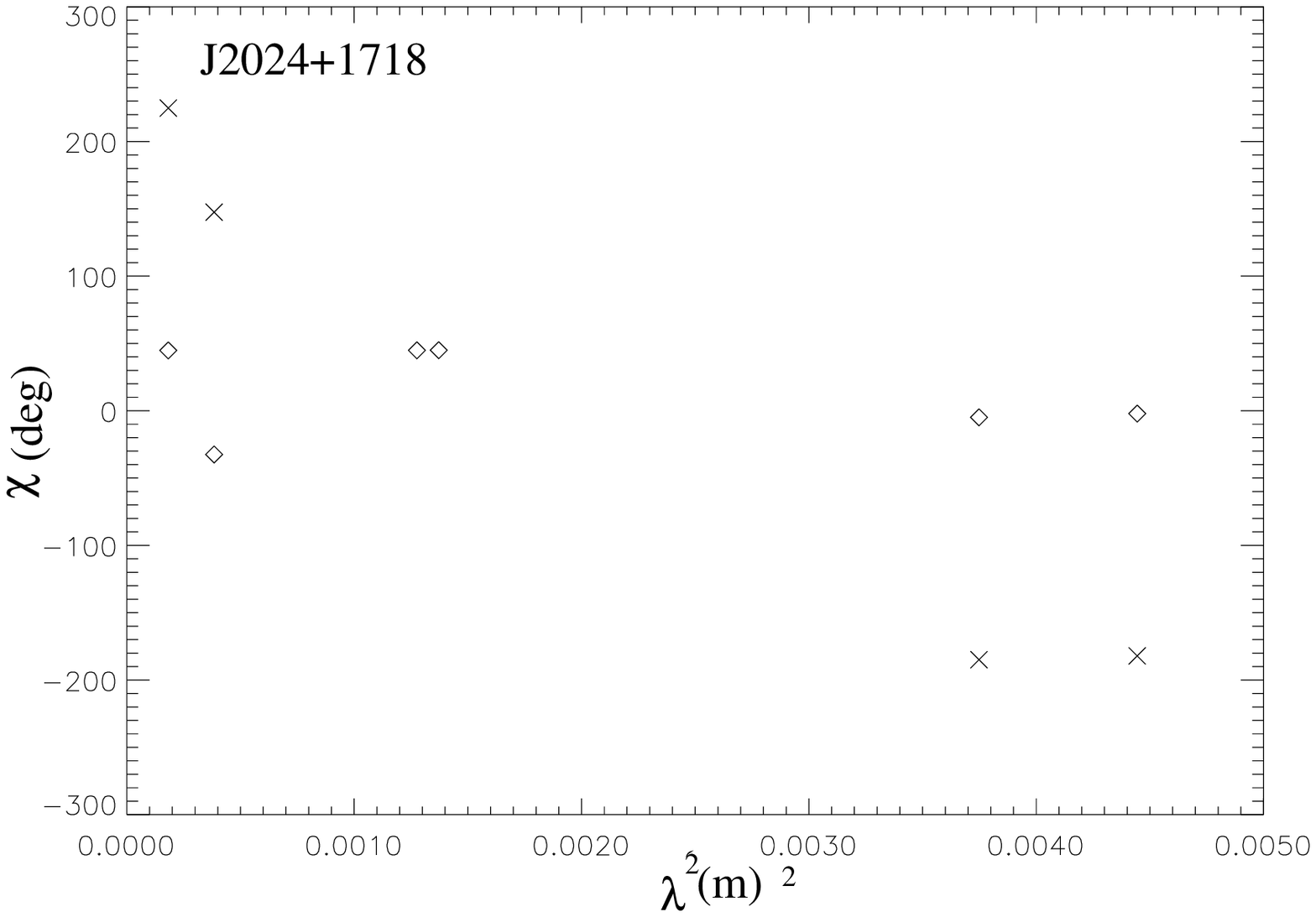}
\vspace{14.5cm}
\caption{Examples of $\chi$ vs $\lambda^{2}$ fits discussed in the
  text. {\it Upper panel:} the fit to the source J2101+0341 as an
  example of a good linear interpolation. {\it Middle panel:} the fit
  to the source J0357+2319 as an example of a good interpolation with
  the addition of $n \pi$. {\it Lower panel:} the case of
  J2024+1718 whose RM cannot be fitted even with the
  addition of $n \pi$.}    
\label{rot}
\end{center}
\end{figure}

In general DP values range between 0 and 1, however in Table \ref{rotodepo}
there are a few sources where $DP > 1$. Depolarization higher
than 1 implies that effects related to the spectral index 
also play a
major role. For example in a source characterized by both an unpolarized
optically-thick core and a polarized optically-thin jet the resulting
DP is greater than 1.\\ 
For the 18 sources ($\sim$40\%) with polarization $>$ 3$\sigma$ 
in three or four bands (providing up to 6 independent data points
given the splitting of the frequencies in C and X bands) we computed
the RM.
In our sample 11 ($\sim 24\%$) sources have $\chi$ measured in four
bands (6 frequencies) and 7 ($\sim 16\%$) 
in three bands (5 frequencies). 
To determine the RM we verify
whether the $\chi$ at different frequencies
are well interpolated by the linear fit. In a few cases we added the
minimum number of $\pm$n$\pi$, such as to have the best least-square
fit to the data (Fig. \ref{rot}). 
In general we find that sources have $\chi$ $\propto$
$\lambda^{2}$. However, for two sources (J0329+3510 and J1616+0459) 
the fit provides high chi-square values, 
while for J2024+1718 no linear fit could
interpolate the data (Fig. \ref{rot}, lowest panel).
As previously mentioned, this may happen if the 
polarized emission originates from two
(unresolved by our VLA data) regions with substantially different RM. \\
If we consider the rotation measure derived between C and K/U bands we
obtain a median value of 74 rad/m$^{2}$, with values as high as
365 rad/m$^{2}$, which are higher than those found in previous works
(e.g. O'Dea \cite{odea89}; Saikia et al. \cite{saikia98}). \\  
For 12 ($\sim 67\%$) of these 18 sources we have information on the
polarization angle at 1.4 GHz from the NVSS (Condon et
al. \cite{condon98}). However, if we consider all polarization angles
from L to K band, we could not find any linear fit across the whole
frequency-range. Here, variability could play a role by changing the
orientation of the E-vector with time. Furthermore, optically-thin
polarized components along the jet on a small scale may provide a
substantial contribution at the lower frequency causing a deviation
from the $\lambda^{2}$ expectations.
If we compute the RM considering the C and L band only, we generally
obtain RM values (median value of 13 rad/m$^{2}$) much lower than
those derived at higher frequency, but this is not significant since
there is no way to solve for $n \pi$ ambiguities. However, it 
also may be
possible that at lower frequency one may be sampling plasma
located at further distances from the core (Saikia et al. \cite{saikia98}).
A KS-test has not found any correlation ($>$99\%) 
between the rotation measure and the
depolarization. \\ 

\begin{table*}
\begin{center}
\caption{Observed and Intrinsic Rotation Measures RM$_{\rm
  obs}$ and RM$_{\rm int}$ computed considering either high-frequency bands
  (from C to K; RM$^{\rm h-f}$), or low-frequency bands (from L and C
  bands; RM$^{\rm l-f}$); 
the Depolarization computed between C and
  X bands, and the intrinsic position angle $\chi_{\rm int}$. 
nf: data not fitted by a linear interpolation. When the redshift
  is unknown, we compute the RM$_{\rm int}$ assuming z=1.0.}
\begin{tabular}{cccccccc}
\hline
\hline
Source&z&RM$_{\rm obs}^{\rm h-f}$&RM$_{\rm int}^{h-f}$&RM$_{\rm obs}^{\rm l-f}$&RM$_{\rm int}^{l-f}$&
DP$_{X}^{C}$&$\chi_{\rm int}$\\
 & &rad/m$^{-2}$&rad/m$^{-2}$&rad/m$^{-2}$&rad/m$^{-2}$& &deg\\
\hline
&&&&&&&\\
J0217+0144&1.715&50&369&20&147&0.6&74\\
J0329+3510&0.5&388&873&12&27&1.7&17\\
J0357+2319& &66&264 &25&100&1.0&-86\\
J0519+0848& &74&296&-39&-156&2.0&69\\
J0625+4440& &-28&-112&5&20&1.1&-80\\
J0642+6758&3.18&-152&-2656&-&-&2.2&-46\\
J0646+4451&3.396&42&811&6&116&0.5&17\\
J1457+0749& &22&88&-&-&1.1&-40\\
J1505+0326&0.411&2&6&43&86&0.2&-46\\
J1616+0459&3.197&202&3558&-&-&0.1&-97\\
J1645+6330&2.379&-129&-1473&-26&-297&0.4&75\\
J1811+1704& &915&3660&-&-&0.8&-58\\
J2024+1718&1.05&nf& &-&-&0.2&\\
J2101+0341&1.013&47&190&-&-&0.9&-75\\
J2123+0535&1.878&373&3090&-13&-108&1.2&-6\\
J2212+2355&1.125&-177&799&19&86&0.8&40\\
J2320+0513&0.622&62&163&-5&-13&1.7&56\\
J2330+3348&1.809&365&2880&-10&-79&2.6&-92\\
&&&&&&&\\
\hline
\end{tabular}
\end{center}
\label{rotodepo}
\end{table*}

\section{Discussion and conclusions}

Multi-frequency polarimetric measurements of a sample of young radio
sources (Fanti et al. \cite{cf04}; Cotton et al. \cite{cotton03})
show that very compact objects ($<$1 kpc) are unpolarized or strongly
depolarized, and the fractional polarization is strictly related to
the frequency: the lower the frequency, the stronger the
depolarization.\\
On the other hand, blazars may not be intrinsically compact, but
they are foreshortened because of projection effects (Antonucci
\cite{anto93}). 
Since these objects are seen at small observing angles, their
  nuclear radio
emission crosses a thinner slab of the magneto-ionic ambient medium
and therefore Faraday rotation and depolarization are only marginally
affected, contrary to what happens in radio sources with the axis 
oriented close to the plane of the sky. 
We find the same behaviour
when we compare the fractional polarization of radio sources with
different optical counterparts. HFPs associated with quasars and BL
Lacs have higher polarization percentages than galaxies and empty
fields, in agreement with unified schemes.\\ 
Several studies of the
polarimetric properties in blazars have shown that these objects
display a fractional polarization from $\sim 1\%$ up to $\sim$10\%, 
almost constant
at all frequencies (Klein et al. \cite{uk03}; Saikia et
al. \cite{saikia98}), 
and a polarization
variability on different time-scales, from a few days to several years
(Saikia \& Salter \cite{ss88}), which can also be unrelated to
the total intensity flux density variability. These objects usually
have small rotation measures, becoming larger moving to higher
frequency (e.g. Saikia et al. \cite{saikia98}; O'Dea
\cite{odea89}). Such a trend also has been found in our sources, 
rotation measures computed between L and C bands being
much smaller than
those between C and K bands. Furthermore, if we consider all the bands
together (from L to K) no linear fit could interpolate the data. 
This may be explained assuming that the low and high frequency
emissions originate from different regions of the radio jet. \\ 
In this scenario, the polarimetric properties are a key element for
the determination of different classes of radio sources.\\
The information derived in this way becomes more effective if other
selection tools, such as flux density variability (Paper I; Tinti et
al. \cite{st05}) and 
morphological information (Orienti et al. \cite{mo06}) 
are taken into account.\\
From the analysis of our multi-frequency polarimetric data, we find 
that 12 sources ($\sim$26\%)
show polarized emission $>$1\%  
at all the available frequencies, 
while another 17 objects ($\sim$38\%) are completely
unpolarized. In a sample of compact sources, such a percentage of
highly-polarized objects reflects a strong contamination by blazar
radio sources.\\  
We find that 14 ($\sim 82\%$) of the unpolarized sources do not show any
significant variability. On the other hand, all the
highly-polarized ($m$ $>$1\%) sources have strong flux-density
variability and 11 ($\sim 65\%$) 
no longer show a convex spectrum.\\
If we analyze the polarized emission in relation to the pc-scale
morphology (Orienti et al. \cite{mo06}), we find that HFPs with or
without a CSO-like structure have different polarization properties:
12 ($\sim 86\%$)
of the CSO-like sources are completely unpolarized at all
frequencies, while 12 ($\sim 60\%$) 
of those without a CSO-like structure have
highly-polarized ($m$ $>$ 1\%) emission at each frequency.\\
All these pieces of evidence confirm the idea that the ``bright'' HFP
sample (Dallacasa et al. \cite{dd00}) is made of two different radio
source populations. If all the discriminant tools (variability,
morphology, polarization) are considered together, we find that 
at least 33 ($\sim 60\%$) objects of the whole sample are
contaminant objects, and only 22 ($\sim 40\%$) 
display all the typical characteristics of young
radio source candidates. Furthermore, all the galaxies of the sample
are still considered young radio sources, supporting the idea
that the majority of galaxies and quasars represent two different
radio source populations.\\

\begin{acknowledgements}
We thank the referee D.J. Saikia for carefully reading the manuscript
and valuable suggestions.
The VLA is operated by the U.S. National Radio Astronomy Observatory
which is a facility of the National Science Foundation operated under
a cooperative agreement by Associated Universities, Inc.
This work has made use of the
NASA/IPAC Extragalactic Database (NED), which is operated by the Jet
Propulsion Laboratory, California Institute of Technology, under
contract with the National Aeronautics and Space Administration.\\
\end{acknowledgements}

\end{document}